\documentclass[11pt]{article}

\usepackage{acl}

\usepackage{times}
\usepackage{latexsym}
\usepackage{booktabs}  
\usepackage{multirow}  
\usepackage{paralist}
\usepackage[T1]{fontenc}
\usepackage{subcaption}

\usepackage[utf8]{inputenc}

\usepackage{microtype}

\usepackage{inconsolata}

\usepackage{graphicx}

\title{Closing the Auto-Research Loop: An AI Co-Scientist for Production Search Ranking}

\author{
  \textbf{Liwei Wu\textsuperscript{1}},
  \textbf{Cho-Jui Hsieh\textsuperscript{2}}
\\
  \textsuperscript{1}Trip.com Group,
  \textsuperscript{2}UCLA
\\
  \small{
    \textbf{Correspondence:} \href{mailto:w@wsw.ai}{w@wsw.ai}, \href{mailto:chohsieh@cs.ucla.edu}{chohsieh@cs.ucla.edu}
  }
}

\begin{document}
\maketitle
\begin{abstract}

We present an AI Co-Scientist framework that closes the research loop for the production search-ranking system of a large online travel platform---pairing LLM agents with direct cloud-compute access so that idea generation, code implementation, GPU experimentation, and result analysis iterate end-to-end with a human scientist in the loop. The framework uses a hybrid agent architecture: single-LLM agents handle routine work, while multi-LLM consensus (GPT-5.2, Gemini Pro 3, Claude Opus 4.5) is invoked for higher-stakes decisions. On the production ranking task, a human-designed transformer baseline (V2) yielded $+0.118\%$ over a pre-transformer baseline (V1); the AI Co-Scientist's automated loop on top of V2 contributed an additional $+0.083\%$, for a combined $+0.201\%$ offline gain delivered in roughly one extra week of wall-clock time (single-run numbers; statistical limits discussed in the paper). The most useful AI proposals---unified long-sequence layouts, slot-type embeddings, and multi-phase learning-rate schedules---are standard practice in NLP and Vision but were absent from our production stack, suggesting that LLM agents can serve as cross-disciplinary connectors for ranking teams. We also report deployment context, negative results, and lessons learned.

\end{abstract}

\section{Introduction}

Recent years have seen rapid progress in AI agents for software engineering and scientific discovery, particularly in drug discovery and materials science. However, applying these general-purpose agents to commercial search ranking remains largely overlooked. While tech giants employ vast teams to manually balance personalization with business objectives, the community has yet to seriously consider using AI agents to assist this process, mirroring broader scientific skepticism about AI's role in guiding research~\cite{gottweis2025towards}. We argue the missing piece is not a smarter single model but a \emph{loop}: an AI co-scientist that, given direct access to cloud compute, can propose, implement, run, and evaluate ranking experiments turn after turn---with a human scientist steering rather than executing.

\begin{figure}[t]
    \centering
    \includegraphics[width=\linewidth]{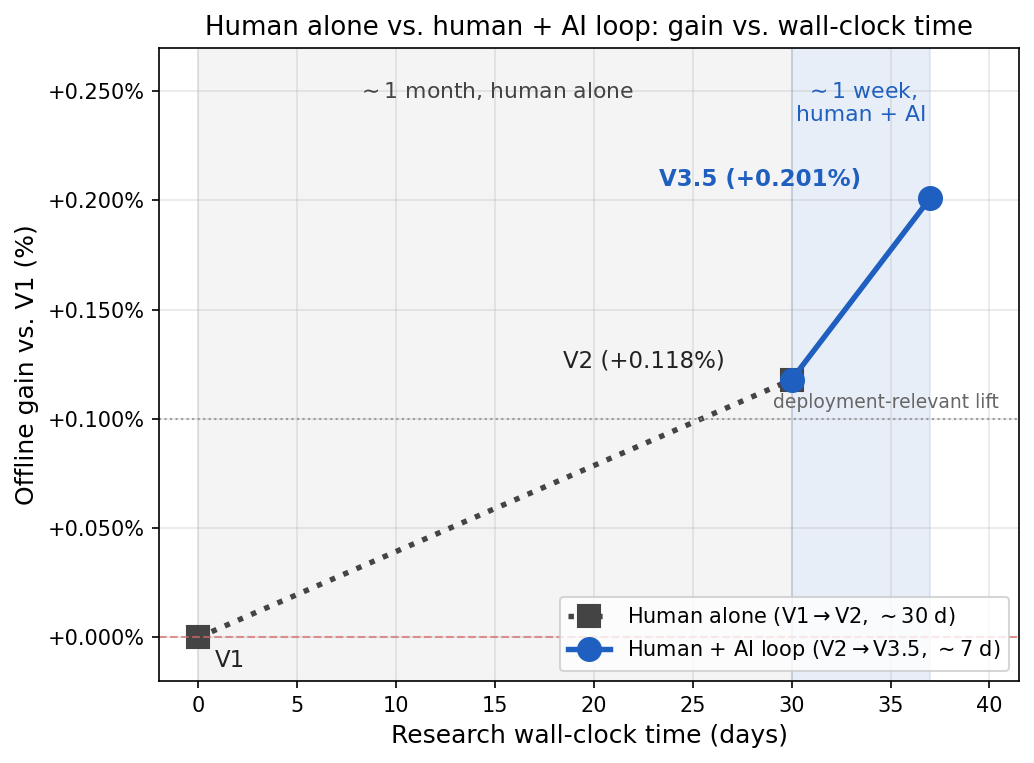}
    \caption{Headline result. Offline gain vs.\ V1 against research wall-clock time: human-alone V1$\to$V2 (${\sim}1$ month, dotted), then human + AI loop V2$\to$V3.5 (${\sim}1$ additional week, solid), combined $+0.201\%$; the compression trades compute for calendar time (Fig.~\ref{fig:performance-gpu-hours}). A single uncontrolled comparison: the dotted segment is schematic and includes infrastructure work the second segment inherited.}
    \label{fig:performance-clock-hours}
\end{figure}

\begin{figure}[ht]
    \centering
    \includegraphics[width=0.88\linewidth]{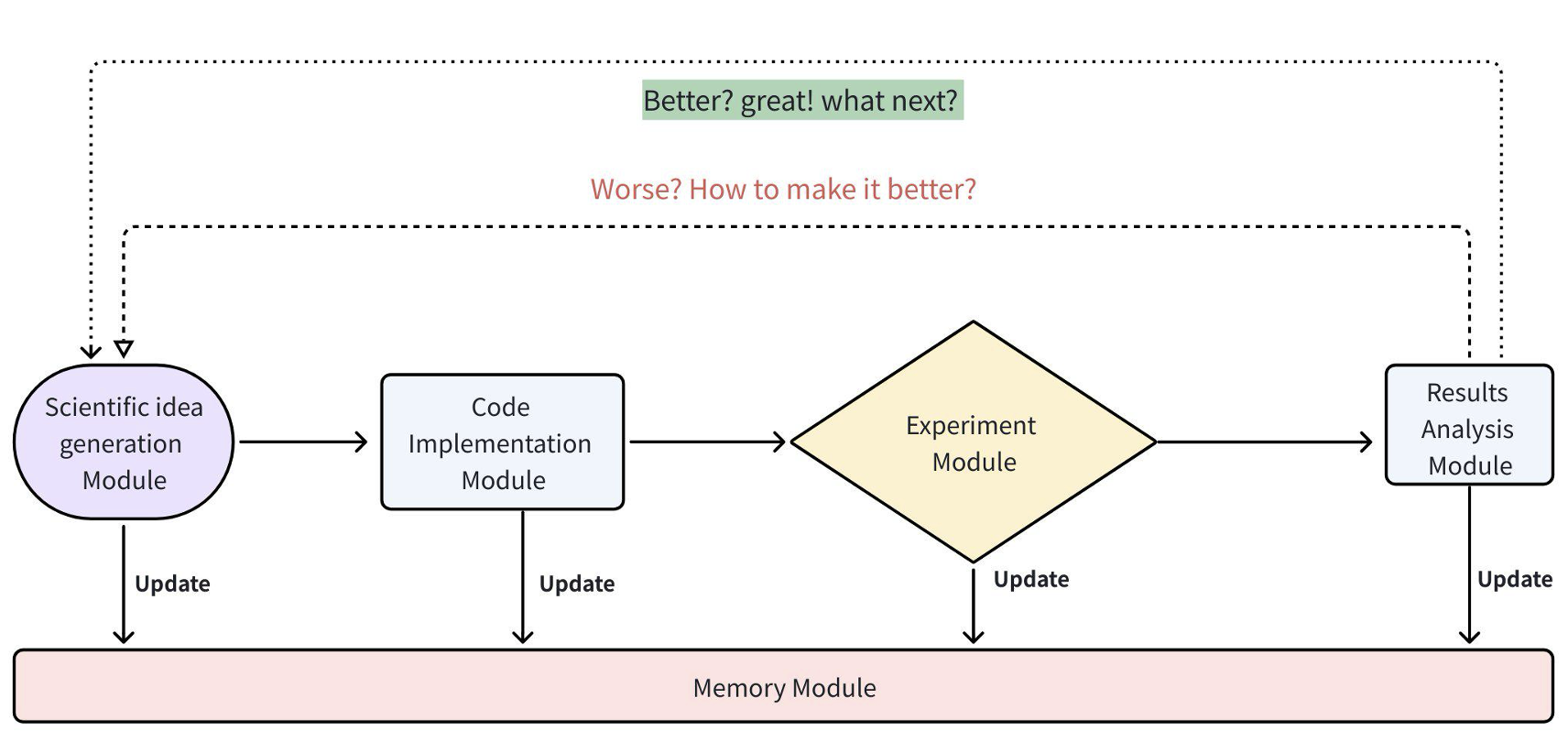}
    \caption{The AI Co-Scientist loop. Idea generation, code implementation, experimentation, and results analysis iterate, each writing state to a shared memory module; the analysis outcome routes the next iteration.}
    \label{fig:ai-agent-design-loop}
\end{figure}

\begin{figure*}[t]
    \centering
    \includegraphics[width=0.7\linewidth]{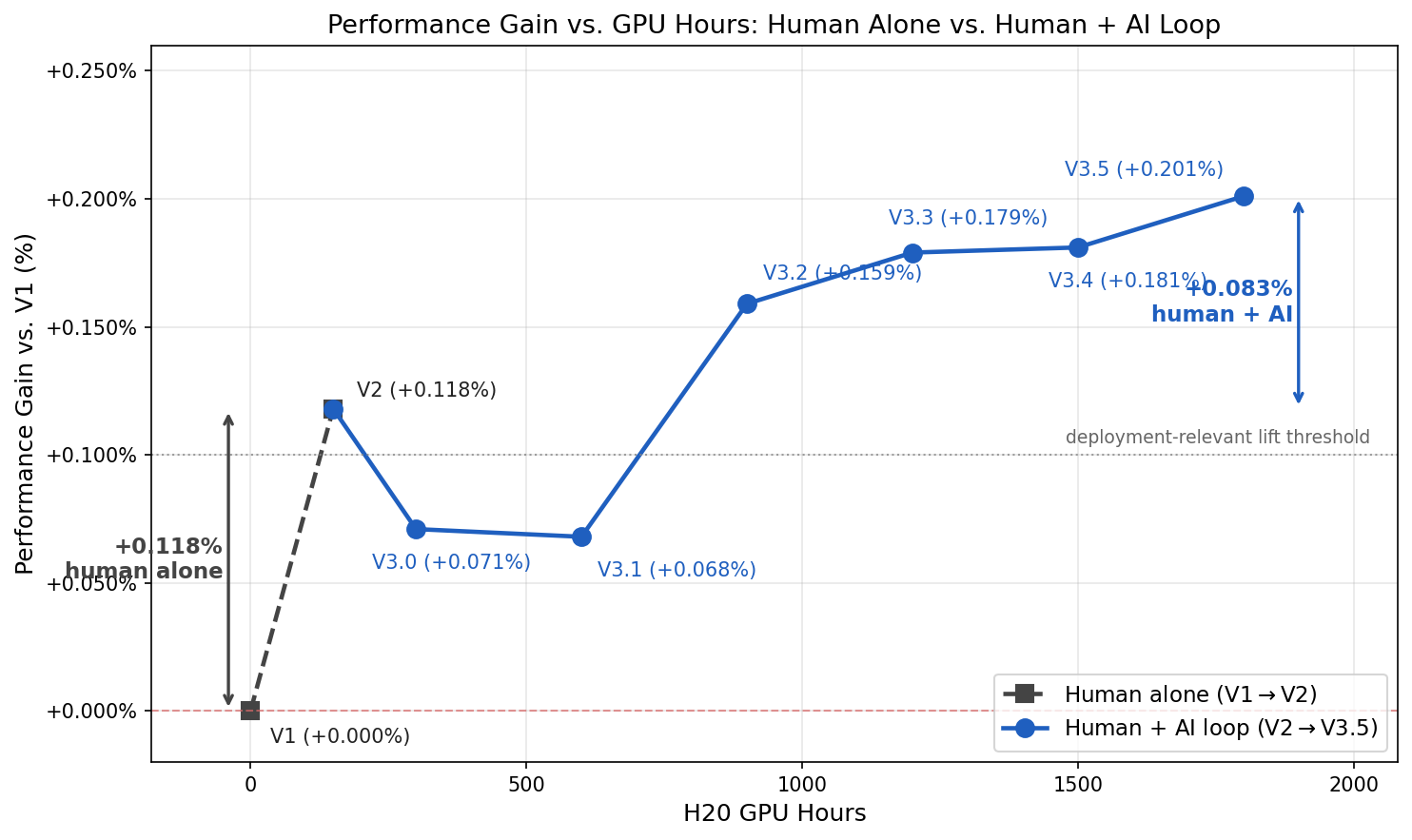}
    \caption{Offline gain vs.\ V1 plotted against H20 GPU-hours: the human-alone V1$\to$V2 segment contributes $+0.118\%$, the human + AI loop V2$\to$V3.5 segment a further $+0.083\%$ on top, combined $+0.201\%$ (V1 ${\approx}0$, V2 ${\approx}150$ are approximate). The dotted reference at $+0.100\%$ is a deployment-relevance threshold (past online experiments at this offline lift translated to ${\sim}0.1\%$ conversion-rate lift), not a statistical-significance bar; the deployment unit is the combined system, which clears it.}
    \label{fig:performance-gpu-hours}
\end{figure*}

In this paper, we instantiate this \emph{auto-research loop} for ranking. By granting LLM-based agents direct access to cloud computing infrastructure, the loop automates the routine phases of the research pipeline---idea generation, code implementation, GPU training-job scheduling, and result analysis---and iterates them under a human scientist's high-level steering, as shown in Fig~\ref{fig:ai-agent-design-loop}. The framework employs single-LLM agents for routine tasks such as code implementation, while invoking multi-LLM consensus agents (GPT-5.2, Gemini Pro 3, and Claude Opus 4.5) for harder decisions including results analysis and idea generation. To the best of our knowledge, this is the first reported study in the ranking community to combine, in a single system, (i) closed-loop iteration over architecture and training schedules, (ii) direct cloud-compute access from the agent rather than human-mediated job submission, (iii) live production data with real opportunity cost, and (iv) a decomposed human-vs-AI gain report that makes each side's contribution falsifiable. Within this loop, all model enhancements from V2 to V3.5 in Table~\ref{tab:model_evolution} (Fig~\ref{fig:architecture-comparison}) were generated automatically and yielded measurable offline gains.

\paragraph{Contributions: a human + machine loop.}
We frame this work as a human + AI auto-research loop rather than autonomous AI research, with three contributions. \emph{First}, we report results as an explicit decomposition: V1$\to$V2 ($+0.118\%$) was human-designed and V2$\to$V3.5 ($+0.083\%$) was generated by the AI Co-Scientist's loop, making the AI's specific contribution visible and falsifiable. The two segments also differ on wall-clock time and compute (Figs.~\ref{fig:performance-clock-hours},~\ref{fig:performance-gpu-hours}): the loop compresses ${\sim}1$ month of research into ${\sim}1$ week by spending roughly an order of magnitude more GPU-hours (${\sim}1{,}650$ vs.\ ${\sim}150$ H20-hours), trading cloud compute for calendar time. \emph{Second}, the agent imported ideas that are standard in NLP and Vision but were absent from our production stack---unified long-sequence layouts, positional and slot-type embeddings, and multi-phase LR schedules (positional encodings initially regressed, an artifact of the old architecture that disappeared under the unified-sequence design; Table~\ref{tab:model_evolution}). Such encodings appear in recent ranking architectures~\cite{zhai2024actions,xu2025climber}; our claim is that the agent surfaced them for our stack unprompted. \emph{Third}, we treat the human scientist's role as a design choice rather than a caveat: the human sets goals, arbitrates consensus disagreements, and intervenes mid-loop on intuition and taste---redirecting the AI along a faster path than blind iteration would find---while the AI absorbs routine experimentation overhead.

\section{Related Work}
\subsection{AI Agents for Scientific Discovery}

The \emph{AI Scientist}~\cite{lu2024ai} has LLM agents autonomously generate ideas, run experiments, and write papers across three ML subfields; we share the end-to-end automation goal but iterate production ranking models rather than papers, with agents tightly integrated with cloud infrastructure. \emph{AI Co-Scientist}~\cite{gottweis2025towards} is a Gemini-based multi-agent system for biomedical hypothesis refinement; we adopt its ``expert-in-the-loop'' paradigm but extend beyond hypothesis generation to code implementation, GPU experimentation, and analysis. AlphaEvolve~\cite{novikov2025alphaevolve} improves algorithms through evolutionary code modification (e.g., improving on Strassen's matrix multiplication). Other domain-specific agents span healthcare~\cite{zhu2025healthflow}, causal discovery~\cite{shen2025exploring}, and mathematics~\cite{romera2024mathematical}.

\paragraph{Comparison to benchmark- and ideation-focused agents.}
Existing AI-agent evaluations in ML mostly fall into two camps: benchmark-driven engineering (e.g.,~\cite{chan2024mle}) on curated tasks in sandboxed environments, and ideation-only studies comparing LLM- vs.\ human-generated ideas. Our setting differs on two axes: (i) the task is a live commercial ranking system with real production data, distribution shift, and a deployment-relevance metric; and (ii) we evaluate end-to-end pipeline execution---ideation, code, training, analysis, and iteration decisions---rather than ideation or engineering in isolation.

\subsection{Scalable Transformer Architecture for Ranking}
Designing novel ranking architectures remains challenging even for experienced engineers. Traditional methods like FM and DCN~\cite{rendle2010factorization, wang2021dcn} are widely used, but optimal transformer design for ranking with sequence features remains open~\cite{zhang2024wukong, zhu2025rankmixer, wang2025scaling}. Recent industrial designs include Climber~\cite{xu2025climber}, OneTrans~\cite{zhang2025onetrans}, HSTU~\cite{zhai2024actions}, and MTGR~\cite{han2025mtgr}, spanning multi-scale sequence extraction, unified feature-interaction/sequence modeling, and trillion-parameter generative recommendation. All are human-designed; to the best of our knowledge, ours is the first reported application of LLM-based agents to this design problem.

\section{Methodology}

Section~\ref{sec:search} briefly formulates the search ranking problem. Section~\ref{sec:co-scientist-ranking} describes the AI Co-Scientist for Ranking, organized into five modules: \begin{inparaenum}[(1)]
    \item Memory (experiment results, code changes, and technical details so far);
    \item Idea Generation;
    \item Code Implementation;
    \item Experimentation (GPU runs);
    \item Results Analysis (interpret results; keep or revert changes).
\end{inparaenum}

\subsection{Search Ranking Problem Formulation}\label{sec:search}

Given a user text query $q$ within scene $s$, the system extracts a $D$-dim dense feature vector $c$ of continuous user/query attributes and $S$ sparse feature sequences $\{f_1, \ldots, f_S\}$, each of length $L$ and sorted oldest-to-latest, capturing different aspects of the user's historical behavior with accommodations. Categorical and real-valued features are bucketized then hashed. Behavior signals form a $J$-dim vector $V$ ($J{=}10$ in practice) of binary or real-valued engagement metrics (e.g., click and conversion). Our goal is to learn a map from $\{c, f_1, \ldots, f_S\}$ to $V$, modeling click-through and conversion probabilities used to rank retrieved accommodations after re-ranking.

\subsection{AI Co-Scientist for Ranking}\label{sec:co-scientist-ranking}

Fig~\ref{fig:ai-agent-design-loop} illustrates the overall process of the AI Co-Scientist for ranking.

\subsubsection{Memory Module}\label{sec:memory}

We adopt a two-layer memory system. The first layer is an index of three files: JOURNEY.md (ideas, results, lessons across iterations), EXPERIMENTS.md (past and active experiments), and FLOWS.md (scheduled training flows). The second layer holds detailed per-module markdown files for each model version, similar to~\cite{cognition2025devin}---e.g.\ V$\{x\}$.$\{y\}$\_IMPLEMENTATION.md records code-implementation details for verification by both AI and human. The AI updates first-layer files after each iteration and only the current version's second-layer files; second-layer files are accessed only when needed.

\subsubsection{Idea Generation Module}\label{sec:idea}
Motivated by recent work~\cite{kasa2025generative}, we suspect that discriminative models can outperform generative models for the search ranking problem because this is a high-noise, high-data, low-model-capacity, high-latency-requirement scenario. We first provide the AI Co-Scientist with the hypothesis that gains observed from earlier transformer-like ranking models come mainly from scaling up the transformer module rather than from a generative next-item loss. We show the AI a worked example (V2's code changes and our manual GPU scheduling scripts) and ask it to propose novel improvements in the same style. Because this module is of central importance and inherently difficult, we leverage multiple state-of-the-art LLMs and have them reach consensus on the next step (Section~\ref{sec:consensus}).

\subsubsection{Multi-LLM Consensus Protocol}\label{sec:consensus}
For the two highest-stakes decisions in the loop---which idea to try next, and whether to keep or revert a code change---we invoke a multi-LLM consensus rather than a single model. Three frontier models run in parallel (GPT-5.2, Gemini Pro 3, Claude Opus 4.5), each prompted independently with the same context (memory state, prior results, candidate options) and asked to (i) propose or evaluate options and (ii) provide a short rationale. We aggregate by majority vote; on direction disagreements the human scientist breaks the tie using the per-model rationales. Single-LLM agents (any one of the three) handle lower-stakes work---code implementation, log parsing, job scheduling---where consensus would be wasteful. Empirically the human arbitrates on a small fraction of calls. The equal-weight vote assumes the three backbones are near-peer models, an assumption indexed to the frontier at the time the experiments ran; we return to this in the Limitations section.

\subsubsection{Code Implementation Module}\label{sec:code}
The AI Co-Scientist implements the code changes corresponding to the idea generated in Section~\ref{sec:idea}. Before writing code, the agent decomposes the change into a dependency graph of routine sub-tasks: setting up the environment, generating boilerplate, and organizing experiment scripts. Explicit task graphs let the human offload low-level coordination while retaining high-level control, and let the planner re-plan affected portions when requirements change. Bugs, when they occur, are typically resolved on a second attempt after reading debug logs.

\subsubsection{Experimentation Module}\label{sec:exp}
In this module, the AI Co-Scientist schedules experiments on GPUs with tunable training parameters after committing the code changes to a new git branch. The same task-decomposition step as in Section~\ref{sec:code} covers configuring hyperparameters, defining evaluation metrics, and scheduling batches.

\subsubsection{Results Analysis Module}\label{sec:results}

We use an aggregated metric $\mathcal{M}$ as the north star for our discovery workflow: the average of 6 AUC metrics across three user behavior types (click, conversion, grouped conversion) and two search scenes (point-of-interest, popularity-based) over 7 days of unseen data. This metric correlates well with online performance and serves as a verifiable reward signal. Throughout the paper we report \emph{deltas} of $\mathcal{M}$ relative to V1 in absolute AUC points $\times\,100$ (e.g., $+0.118$ corresponds to $+0.00118$ in raw AUC).
The AI Co-Scientist computes $\Delta\mathcal{M}$ against the previous best and proceeds accordingly: if $\Delta\mathcal{M} > 0$, it creates a new branch and implements the next iteration; if $\Delta\mathcal{M} < 0$, it decides whether to fix the current branch or revert to the previous best. This decision is controlled by a threshold parameter---higher values encourage persistence, lower values favor quick reversion. As in Section~\ref{sec:idea}, multiple LLMs reach consensus on this decision via the protocol of Section~\ref{sec:consensus}.

\begin{figure}[ht]
    \centering
    \begin{subfigure}{0.37\textwidth}
        \centering
        \includegraphics[width=\linewidth]{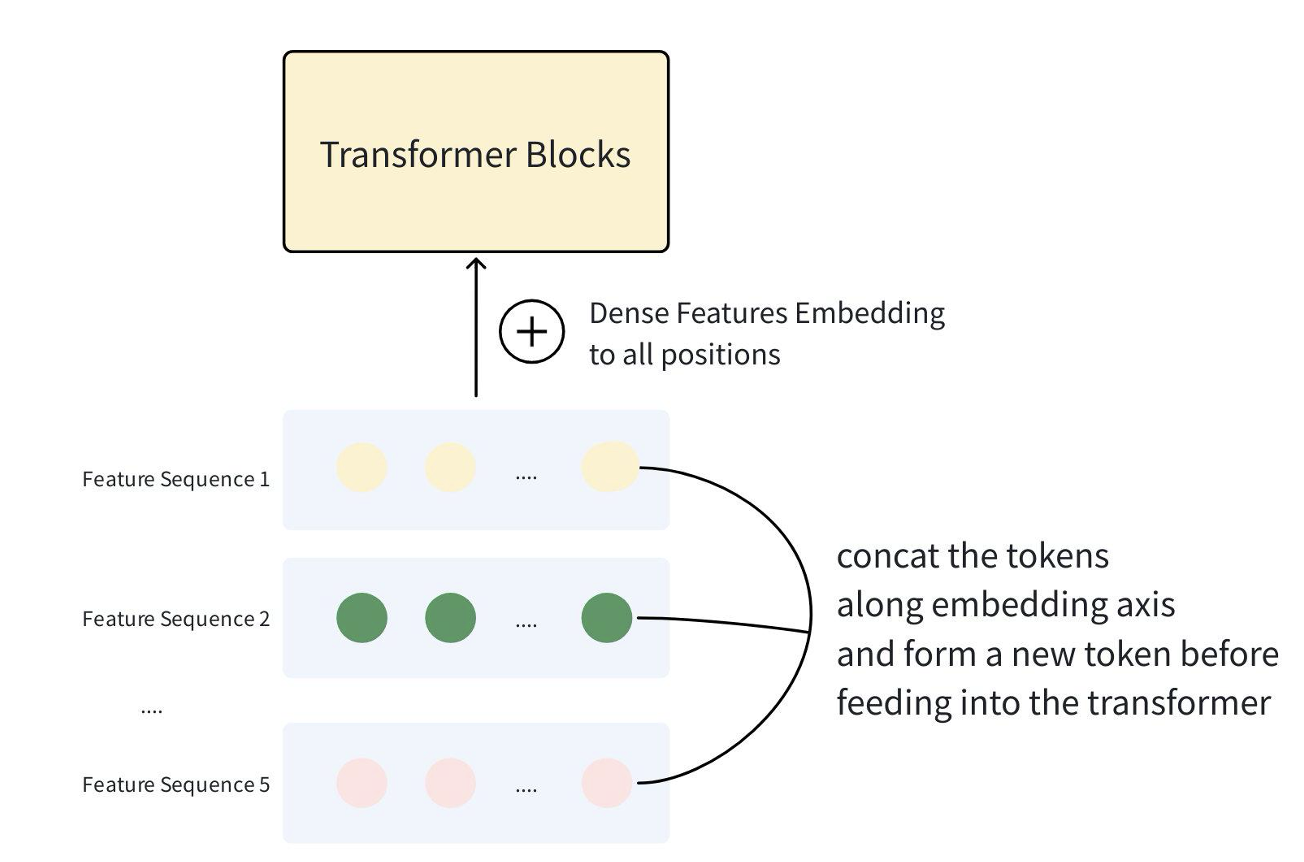}
        \caption{V2 design}
        \label{fig:v2}
    \end{subfigure}
    \hfill
    \begin{subfigure}{0.37\textwidth}
        \centering
        \includegraphics[width=\linewidth]{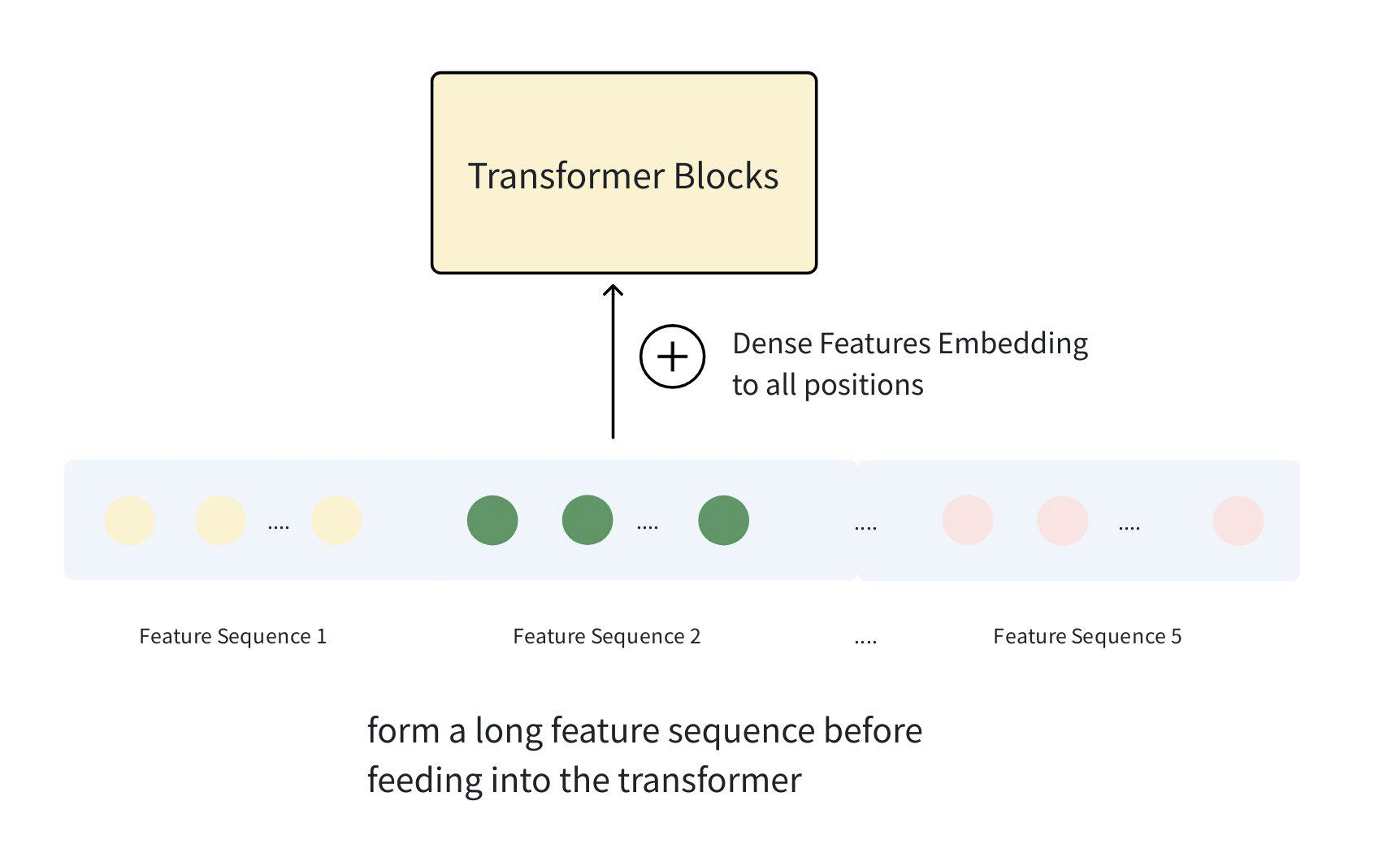}
        \caption{V3 design}
        \label{fig:v3}
    \end{subfigure}
    \caption{Comparison of Transformer designs}
    \label{fig:architecture-comparison}
\end{figure}

\begin{table*}[htbp]

\centering

\resizebox{0.8844\textwidth}{!}{%

\begin{tabular}{llcccc}

\toprule

\textbf{Version} & \textbf{Key Innovations} & \textbf{Seq Len} & \textbf{LR} & $\mathcal{M}$ (\%) \textbf{vs V1} & \textbf{Status} \\

\midrule

V1 & Mean-Pooling + DCN + MOE & -- & $5\times10^{-5}$ & $0.000$ & Baseline (pre-transformer) \\

\midrule

V2 & Transformer + Separate Sequences (Figure~\ref{fig:v2}) & 40 & $5\times10^{-5}$ & $+0.118$ & Human-designed baseline \\

\midrule

V3.0 & V2 + Positional Encoding + Attention Pooling & 40 & $5\times10^{-5}$ & $+0.071$ & Regressed \\

V3.1 & Transformer + Unified Sequence (Figure~\ref{fig:v3}) & 200 & $5\times10^{-5}$ & $+0.068$ & Regressed (LR diagnosed in V3.2) \\

V3.2 & V3.1 + Reduced LR & 200 & $1\times10^{-5}$ & $+0.159$ & Improved \\

V3.3 & V3.2 + Slot Type Embeddings & 200 & $1\times10^{-5}$ & $+0.179$ & Improved \\

V3.4 & V3.3 + Temporal Embeddings & 200 & $1\times10^{-5}$ & $+0.181$ & Marginal (within run noise) \\

\textbf{V3.5} & \textbf{V3.4 + Four-phase LR optimization} (Table~\ref{tab:four_phase_lr}) & \textbf{200} & \textbf{Adaptive} & $\textbf{+0.201}$ & \textbf{New Best (AI-generated)} \\

\bottomrule

\end{tabular}%

}

\caption{Transformer model evolution: architecture and offline performance vs.\ V1. V1$\to$V2 was human-designed (+0.118\%); V2$\to$V3.5 was generated by the AI Co-Scientist (+0.083\%); the combined gain is +0.201\%. All entries are single best runs without repeated seeds or confidence intervals; in particular, the V3.3$\to$V3.4 step (+0.002) is too small to distinguish from run-to-run variation.}

\label{tab:model_evolution}

\end{table*}

\section{Experiments and Findings}

\subsection{Generating and Evaluating Ranking Architectures}
We treat the use of only five of the more than 200 available feature-sequence slots as a pilot study, feeding only these five into the transformer. V1 is a pre-transformer baseline that mean-pools sequence features before DCN~\cite{wang2021dcn}. V2 is human-designed: it stacks sequence features and projects dense features into an embedding added to sequence embeddings at all positions, following standard practice in personalized transformers~\cite{wu2020sse} and similar to~\cite{xu2025climber}. V2 alone yields $+0.118\%$ over V1 and forms the launch point for AI-generated iterations. All numbers in this section are single best runs without repeated seeds or confidence intervals; we flag below where differences are too small to interpret. Building on V2 (Fig.~\ref{fig:v2}), the AI proposes V3 (Fig.~\ref{fig:v3}). As Table~\ref{tab:model_evolution} summarizes, V3.0 (positional encodings + attention pooling) regresses against V2; V3.1 concatenates the separate sequences into one long sequence and also underperforms; V3.2 identifies the cause as too-large LR and reduces it to one fifth after half the data, reaching $+0.159\%$.

\subsection{Designing Learning-Rate Schedules}
Training convergence is challenging for multi-task ranking models~\cite{tang2023improving}, and tuning learning rates is hard even for human experts. We have the AI Co-Scientist design LR schedules from observed training-dynamics data. It eventually proposes V3.5 (Table~\ref{tab:four_phase_lr}): standard LR, then a one-fifth reduction, then a brief increase to escape a plateau, then a final reduction for fine-tuning. V3.5 reaches $+0.201\%$ over V1. We note that the V3.3$\to$V3.4 step ($+0.002$) is too small to distinguish from run-to-run variation without repeated seeds; we report it for provenance, not as an established gain.

\begin{table}[htbp]
\centering
\caption{V3.5 four-phase learning rate optimization strategy.}
\label{tab:four_phase_lr}
\resizebox{\columnwidth}{!}{%
\begin{tabular}{lccl}
\toprule
\textbf{Phase} & \textbf{Days} & \textbf{Learning Rate} & \textbf{Purpose} \\
\midrule
0  & 1--16 & $5\times10^{-5}$ & Standard LR \\
 1 & 17--19 & $1\times10^{-5}$ & Reduced LR\\
 2 & 20--23 & $1.5\times10^{-5}$ & Plateau Escape \\
 3 & 24--28 & $8\times10^{-6}$ & Final Fine-Tuning \\
\bottomrule
\end{tabular}%
}
\end{table}

\paragraph{On the $0.1\%$ threshold and statistical validity.}
The deployment unit is the combined human + AI system: the loop is an additive layer on top of human work, and the joint $+0.201\%$ clears the deployment-relevance bar. The $0.1\%$ marker in Figure~\ref{fig:performance-gpu-hours} is not a significance bar but a deployment-relevance threshold (offline lifts of this size have historically translated to ${\sim}0.1\%$ online conversion lift); the $+0.083\%$ AI-only slice is reported for provenance, not as a standalone deployment claim. Multi-seed runs and confidence intervals remain a clear next step.

\paragraph{On serving cost.}
Attention cost is quadratic in sequence length, so a unified length-200 sequence costs roughly $5\times$ the attention FLOPs of five separate length-40 sequences at equal width; offline gains become deployment candidates only if the production latency budget is also met.

\paragraph{On the compute-substitution hypothesis.}
A natural counter is that a human given the same ${\sim}1{,}650$ H20-hours and an extra week might match the V2$\to$V3.5 gain. We cannot refute this without a parallel human-only arm, but two features make it hard to sustain. First, the GPU-hours were spread across many short branches (Table~\ref{tab:model_evolution}), several of which initially regressed; serializing that trajectory by hand costs \emph{calendar} time to read each result and choose the next branch---precisely the bottleneck the loop removes. Second, the gain-producing ideas are documented imports from NLP and Vision rather than ranking defaults; the substitution argument must also claim the human would have proposed the same imports on the same timeline, which the V1$\to$V2 segment does not support.

\paragraph{On not ablating across LLM backbones.}
We do not re-run V2$\to$V3.5 with weaker LLM backbones: the system runs against live production data with real opportunity cost, so an ablation arm would consume GPU and human-review budget comparable to the original run. The contribution we report is the \emph{loop architecture}---memory, multi-LLM consensus, planning, and analysis---rather than a capability score for any single backbone; this is a methodological stance, not a claim that backbone choice is irrelevant.

\subsection{Deployment Status and Lessons Learned}\label{sec:lessons}

\paragraph{Deployment status.}
The V2$\to$V3.5 stack subsequently entered the platform's standard release process, in which an online A/B test bundles many concurrent changes; per-change online attribution is therefore not possible, and we make no online claim for V3.5 in isolation. This is an industry reality rather than an omission: changes are promoted through an offline gate and verified online at bundle granularity, so the offline decomposition we report is the attribution the production process actually produces.

\paragraph{Lesson 1: a verifiable reward is what closes the loop.}
The single aggregated AUC-delta $\mathcal{M}$ (Section~\ref{sec:results}) did more work than any agent component: it made keep-or-revert decisions checkable by both the AI and the human, and it is the reason the loop could run without per-step human sign-off.

\paragraph{Lesson 2: cheap iteration changes how regressions are read.}
V3.0 and V3.1 looked like architecture failures. The loop's low iteration cost made it affordable to test the alternative hypothesis that the inherited learning rate was at fault, which V3.2 confirmed ($+0.068\%\to+0.159\%$); a team paying full calendar cost per experiment plausibly reverts the architecture instead.

\paragraph{Lesson 3: metric-winning bugs make code audit non-optional.}
The agent occasionally produced implementations that improved $\mathcal{M}$ while being wrong (e.g., silently increasing model size). Human-legible memory files (the per-version implementation notes of Section~\ref{sec:memory}) are the audit surface that makes such review possible; the Limitations section discusses the residual risk.

\paragraph{Lesson 4: spend consensus only where it pays.}
Multi-LLM consensus was reserved for the two highest-stakes calls---idea selection and keep-or-revert---while single-LLM agents handled code, log parsing, and scheduling. We saw no benefit that would justify paying the $3\times$ cost on routine work.

\section{Conclusion}

A recent interview-based report \cite{anthropic2025interviewer} found that AI is still unable to manage central aspects of scientific work such as formulating hypotheses and conducting experiments. Our findings suggest that, on at least one production-ranking task, an AI Co-Scientist paired with a human scientist and given cloud-compute access can take on a non-trivial share of the routine research workload, contributing $+0.083\%$ on top of a human-designed $+0.118\%$ V2 baseline for a combined $+0.201\%$. We frame this as human + machine collaboration, not autonomous AI research, and view freed-up researcher time as input to higher-level activities rather than a replacement for them. The loop's working parts are ordinary modules wired in a specific way---two-layer memory, consensus reserved for high-stakes calls, task decomposition, a single verifiable reward (Section~\ref{sec:lessons})---and the assembly, not any one module, is what we expect to transfer beyond ranking. Generalizing beyond a single ranking task is left to future work.

\paragraph{Hypothesis: AI as a cross-disciplinary connector.} The AI Co-Scientist's most useful proposals are standard practice in NLP and Vision but were absent from our production stack. Ranking teams understandably read most deeply within their own subfield; an LLM agent has no such habit and surfaces primitives from adjacent ML areas at no social cost. The same role may extend to non-adjacent disciplines---dynamical-systems formulations from physics, population-level objectives from evolutionary biology. Whether far-field transfers yield offline gains, and how to keep the search productive rather than eclectic, is left for future work.

\section*{Limitations}

We treat the limitations below as first-class. The first four are concrete properties of the system reported here; the last three concern longer-horizon and conceptual risks that this paper inherits but does not directly evaluate.

\paragraph{Code reliability and silent bugs.}
The AI agent occasionally introduces bugs in code implementations that can be difficult to detect without expert supervision. For instance, it may unintentionally increase model size or introduce subtle errors in architecture design code. In some cases, judged purely by experimental metrics, these buggy implementations can appear to ``win'' and be selected as a better model even though the underlying code is incorrect.

\paragraph{Inefficient idea generation and resource usage.}
Idea generation by the AI agent is still not very efficient and can lead to substantial waste of GPU resources. A human scientist typically remains in the loop to prevent catastrophic failures (e.g., obviously broken configurations or experiments that do not meaningfully contribute to progress).

\paragraph{Myopic feedback loop.}
The optimization loop is vulnerable to myopic behavior: the agent may favor ideas yielding short-term improvements in experimental metrics that are not promising in the long run. Such short-term gains can mislead the agent and reinforce directions a human scientist would deem uninteresting.

\paragraph{Knowledge base integration.} The system currently lacks access to scientific databases and real-time literature search, limiting its use of state-of-the-art methods.

\paragraph{Consensus under an uneven frontier.}
The equal-weight majority vote of Section~\ref{sec:consensus} implicitly assumes its three backbones (GPT-5.2, Gemini Pro 3, Claude Opus 4.5) are near-peer models, which was a reasonable reading of the frontier when our experiments ran. It may not hold at replication time: with the release of Claude Fable 5, now widely regarded as stronger than other frontier models, an equal-weight vote lets two weaker models outvote the strongest one on exactly the high-stakes decisions the protocol was built for. The protocol should therefore be read as indexed to the frontier balance at run time; capability-weighted voting, or a strongest-model-as-leader design with weaker models as critics, are natural adjustments we have not evaluated. Replication also depends on access: regimes that gate frontier models by nationality or geography partition the research community and undercut the cross-backbone reproduction this protocol assumes---equitable cross-border access is, in our view, a precondition for reproducible open research.

\paragraph{Role of research taste.}
Human research taste remains important: it can guide the process toward more elegant and principled ideas and designs even when those do not yield immediate experimental gains, and without such guidance the system may overemphasize short-term improvements and miss promising long-term directions. We note, however, that this need not be a permanent division of labor. Even on a conservative reading---in which AI never acquires real research taste---humans spend most of their time on the small, direction-setting fraction of work while AI handles the rest, so each researcher steers far more work than before~\cite{favaro2026rsi}. A less conservative reading is that taste is itself another capability that AI systems fail at for a time and then get good at, following the same pattern as theory of mind, explaining humor, and linguistic puzzles~\cite{favaro2026rsi}; under that reading, the human-taste assumption underpinning our current loop should be expected to weaken over time rather than hold fixed.

\paragraph{Recursive self-improvement risk in autonomous research loops.} A distinct, longer-horizon concern is that autonomous research loops---in which an AI system proposes ideas, implements them, evaluates them, and feeds the results back into the next round---are an early, narrow instance of the recursive self-improvement dynamic recently flagged by Favaro and Clark~\cite{favaro2026rsi}: as ``the doing'' becomes nearly free in human time, human review becomes the bottleneck, and rare misalignments in today's models risk compounding through successive iterations until oversight is lost. Our system mitigates this in the small by keeping a human scientist on the design and selection path and by limiting the action space to ranking-model code, but the failure mode generalizes: any production auto-research stack that closes the loop on its own outputs, especially one that begins to modify the agent or its evaluation harness, inherits this risk and demands oversight, validation, and verification mechanisms beyond what we exercise here.

\paragraph{AI safety considerations.} While the system only generates ranking models, AI safety remains important for production deployment. Future work should add automated guardrails beyond the current reliance on human scientists.

\section*{AI Use Disclosure}

Fittingly, this revision was itself a human+AI collaboration: the authors set the argument, claims, and final wording, while Claude Code (Anthropic) assisted with LaTeX edits, figure code, and prose polishing. All technical claims, results, and analysis are the authors'.

\bibliography{acl_latex}


\end{document}